\documentclass[10pt,a4paper]{article}

\usepackage[T1]{fontenc} 

\usepackage[english]{babel} 

\usepackage{abstract} 

\usepackage{titlesec} 
\renewcommand\thesection{\Roman{section}} 
\renewcommand\thesubsection{\roman{subsection}} 
\titleformat{\section}[block]{\large\scshape\centering}{\thesection.}{1em}{} 
\titleformat{\subsection}[block]{\large}{\thesubsection.}{1em}{} 

\usepackage{titling} 

\usepackage{hyperref} 
\usepackage{natbib}
\usepackage{graphicx}

\title{\LARGE \bf Methodology to create a new Total Solar Irradiance record: Making a composite out of multiple data records}
\author{Thierry Dudok de Wit$^1$, Greg  Kopp$^{2,3}$, Claus Fr\"ohlich$^4$, Micha Sch\"oll$^{1,5}$ \\[1ex] 
\small $^1$ LPC2E, CNRS and University of Orl\'eans, France \\ 
\small $^2$ Laboratory for Atmospheric and Space Physics, University of Colorado, Boulder, CO, USA \\ 
\small $^3$ Max-Planck-Institut f\"ur Sonnensystemforschung, G\"ottingen, Germany \\
\small $^4$ D\"ahlenwaldstrasse 30, Davos Wolfgang, Switzerland \\ 
\small $^5$ Physikalisch Meteorologisches Observatorium Davos and World Radiation Center, Davos Dorf, Switzerland \\ 
}

\renewcommand{\maketitlehookd}{%
\begin{abstract}
\noindent Many observational records critically rely on our ability to merge different (and not necessarily overlapping) observations into a single composite. We provide a novel and fully-traceable approach for doing so, which relies on a multi-scale maximum likelihood estimator. This approach overcomes the problem of data gaps in a natural way and uses data-driven estimates of the uncertainties. We apply it to the total solar irradiance (TSI) composite, which is currently being revised and is critical to our understanding of solar radiative forcing. While the final composite is pending decisions on what corrections to apply to the original observations, we find that the new composite is in closest agreement with the PMOD composite and the NRLTSI2 model. In addition, we evaluate long-term uncertainties in the TSI, which reveal a $1/f$ scaling. 
\end{abstract}
}

\date{\normalsize Slightly expanded version of a manuscript published in Geophysical Research Letters (2017), \url{http://dx.doi.org/10.1002/2016GL071866}} 

\begin{document}

\maketitle

\section{Introduction}

Combining different (and only partly overlapping) time series of the same physical quantity into a single composite is both a scientific and a statistical challenge that arises in many contexts, in particular in paleoclimatic reconstructions \citep{mann08}. In space sciences, observations are often constrained by the finite lifetimes of satellites, making composites the key to investigation over long timescales. A timely and demanding application is the reconstruction of the total solar irradiance (TSI), which is the spatially- and spectrally-integrated radiant output from the Sun at a mean Sun-Earth distance of 1 astronomical unit \citep{kopp14}. 

The TSI has been continuously measured since November 1978 by over a dozen instruments, and is paramount to  understanding the Earth's global energy budget \citep{trenberth09}.  Weak secular variations in the TSI are hotly debated, as they may have large implications on our understanding of the role of the Sun in climate change \citep{ermolli13}. 

The nominal value of the TSI, averaged over Solar Cycle 23 (which lasted from 1996 to 2008), is $1361.0\pm 0.5$ [W/m$^2$], with a weak peak-to-peak solar-cycle modulation of 0.08\% that is in phase with the 11-year cycle \citep{kopp16}. Assessing such tiny modulations requires not only high radiometric accuracy but also considerable care in the making of the composite record. 

There currently exist three commonly-used TSI composites  \citep{willson97,froehlich04,mekaoui08}, all of which are made by daisy-chaining: different records are stitched together by comparing them during an overlap period when at least two instruments are observing \citep[e.g.][]{froehlich97a}. This approach has several shortcomings. For each day, only a single instrument is selected for building the composite, and thus complementary cotemporal information is lost. In addition, the choice of the most trustworthy instrument introduces a bias toward preconceived ideas of how the TSI should vary.

Largely because of these shortcomings, these three composites show different trends in time \citep{zacharias14}, which has fueled a continuing debate. A trend in the TSI measurement record, if any, would have major implications on the historical reconstruction of TSI from models. To address this issue, the decision was made to create a new composite that would be fully traceable and based on community involvement. This composite should combine the original records with no reliance on any external proxy by using state-of-the-art statistical methods. Here, we concentrate on this methodology and introduce a novel, probabilistic approach that can be readily exported to other contexts.



\begin{figure}[hbt]
\begin{center}
\noindent\includegraphics[width=0.95\textwidth]{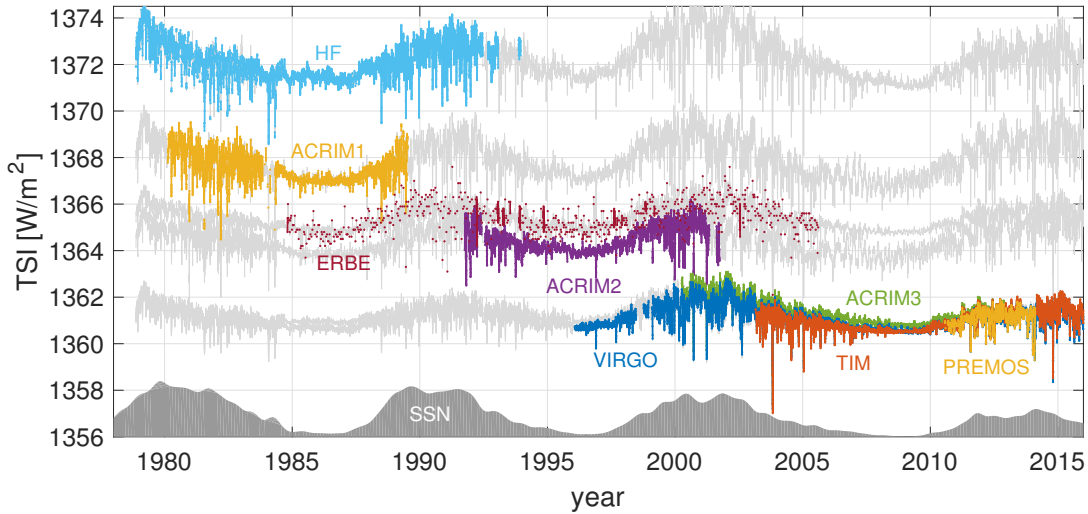}
\caption{Original data sets used in this study, in absolute units. Also shown is the sunspot number (SSN), averaged over 6 months, in arbitrary units. The gray extension of each record represents the reconstruction obtained by expectation-maximization, see Sec.~\ref{sec_methodology}. These extrapolations are used for the sole purpose of applying the wavelet transform in a systematic way and are not used for analysis purposes. }
\label{fig_original_data}
\end{center}
\end{figure}

\section{Original data}

The eight instruments that are routinely used for making the TSI composite are listed in Table 1 and their records are illustrated in Fig.~\ref{fig_original_data}. Most observe on a daily basis, with occasional interruptions and outliers. Usually, one to three of them are operating simultaneously, although some days are devoid of observations. Figure~\ref{fig_original_data} highlights the challenge of measuring a weakly-varying TSI in absolute units: all instruments agree well on the amplitude of relative variations but differ in their baseline. The two most recent radiometers (TIM and PREMOS) are the only ones that accounted for internal-instrument scatter effects prior to launch, and thus do not require subsequent large corrections as the others do. These two instruments will hereafter serve as a reference for the absolute value of the TSI for this paper.

Our main working hypotheses are that all observations are 1) well dated and resolve daily variations of the TSI, 2) made on a regular time grid, which prevents needing to resample them, and 3) correct except for an offset in their absolute value and errors that are within the uncertainty. Short gaps in observations and uncorrected instrument drifts are also acceptable.

Daily TSI observations actually mask disparities: all instruments make several observations per day (with a cadence of up to 50 seconds for TIM) and then average these to produce a daily average. ERBE, however, on average observes the Sun once every 14 days for 3 minutes.  Instruments also differ in the way their in-flight degradation is corrected, and several records suffer from issues such as occasional satellite off-pointing. Such effects emphasize the need for including time-dependent uncertainties in the composite; these are lacking in present composites.

\begin{table}
\begin{center}
\begin{tabular}{|l|c|c|c|}

\hline
 Instrument & Version  & Start date  & End date   \\
\hline
HF / NIMBUS-7 ERB	&		& 1978/11 	& 1993/01 		 \\
ACRIM1 / SMM	 	&	a	& 1980/02 	& 1989/07 		\\
ERBE / ERBS			&   	& 1984/10	& 2003/08 		 \\
ACRIM2 / UARS		& 	b	& 1991/10	& 2000/09 			\\
VIRGO / SOHO		&   c	& 1996/01	& active 		\\
ACRIM3 / ACRIMSAT \ \ \	&   d   & 2000/04	& 2013/02 			\\
TIM / SORCE			&   e   & 2003/03	& active 		\\
PREMOS / PICARD   	&   f   & 2010/07	& 2014/02 			\\
\hline
\multicolumn{4}{l}{Version: a) 1, b) 7/14, c) 6.005.1602, d) 11/13, e) 17, f) 1} \\

\end{tabular}
\caption{Observations used in this study. For details on the individual instruments, see \citet{froehlich12}. All instruments provide daily values except for ERBE. Some instruments that are less frequently used and/or have been observing for shorter periods are not included in this list.}
\end{center}
\end{table}


\section{Methodology}
\label{sec_methodology}

Raw TSI data are displayed in Figure~\ref{fig_original_data}, showing that few instruments have observed the TSI for more than a solar cycle.  One of the greatest challenges with daisy-chaining is maintaining continuity across data gaps. In particular, the impact of the two-year data gap from July 1989 to October 1991 from when ACRIM1 ceased operating to when ACRIM2 started remains hotly debated because the only instrument observing during this time, the ERBE, had no degradation-correction monitor of its single sensor. 

The probabilistic approach we advocate overcomes these problems in a natural way with two significant improvements over existing approaches. First, we use all available observations, weighted by their uncertainties. Ideally, we would rely on time-dependent uncertainties provided by the instrument teams. However, these are rare and, even when available, often cannot be meaningfully compared because they do not use a common estimation-methodology.  At best, they may serve in relative terms, for example to indicate how the uncertainty evolved during mission life. To bypass that issue and compare the instruments more uniformily, we require a common metric for all data sets. For that purpose, we use a data-driven approach, which does not rely on any preconceived proxy or TSI model. Second, we decompose the data into different timescales and perform the averaging scale-wise. While instruments may agree well on one timescale (e.g., the 27-day solar rotation period), they may give more contrasting results at other scales.


\section{Uncertainty Estimation}

Uncertainty of the TSI can be classified as precision, stability, or accuracy \citep{schoell16}.  Precision is the error associated with random fluctuations and can be estimated by various means \citep{ddw16}. Estimating the stability, which is the error associated with long-term variations, is considerably more difficult. Finally, accuracy concerns the error on the absolute value, which is determined by instrument calibrations. Accuracy estimation is beyond the scope of this study, so we use as an absolute value the average of TIM and PREMOS, which agree with only small differences that are well within their estimated accuracy uncertainties. This value is also consistent with that reported by \citet{kopp11}.

Precision is usually estimated by considering the high-frequency components of the signal, which we assume to be a mix of a slowly-varying solar signal with additive incoherent fluctuations that are mostly of instrumental origin. This is indeed supported by those instruments that observe with sub-daily cadence. To estimate the fluctuation level we consider an autoregressive model \citep{Mann96,ddw16}, which provides an estimate $\hat{I}(t)$ of the present value  of the TSI from a linear combination of its $p$ previous observations
\begin{equation}
\label{eq_AR_model}
\hat{I}(t) = a_1 I(t-1)  + \ldots + a_p I(t-p) \ ,
\end{equation}
where $I(t)$ is the TSI variability relative to its time-averaged value. The difference $\epsilon(t) = I(t) - \hat{I}(t)$ between the observed and modeled TSI, which is called innovation, represents the non-reproducible high-frequency noise. We find that models of order $p \approx 5$ provide a good compromise between goodness of fit and overfitting, with no substantial decrease in the innovation for larger orders (see Appendix 1). In what follows we thus set $p=5$ and fit the model independently to each record. Our precision is now given by $\sigma_I = \sqrt{\langle \epsilon(t)^2 \rangle_t}$. For this particular study, we consider $\sigma_I$ as constant in time, although one could easily let it vary in time. The average precision ranges from 0.07 [W/m$^2$] (for TIM) to 0.8 [W/m$^2$] (for ERBE). For TIM and PREMOS, these values are within a factor of two of the precisions stated by the instrument teams. For ACRIM2 and older instruments, the precisions are systematically larger than stated.

The quantity of prime interest here is the uncertainty on longer timescales, i.e. the stability.  Estimating it without the help of any external reference is a notoriously difficult task. Most studies implicitly make a white noise hypothesis, which is akin to saying that stability equals precision.  While this assumption is mathematically convenient, it is unproven for existing TSI instruments. 

\begin{figure}[!ht]
\begin{center}
\noindent\includegraphics[width=0.8\textwidth]{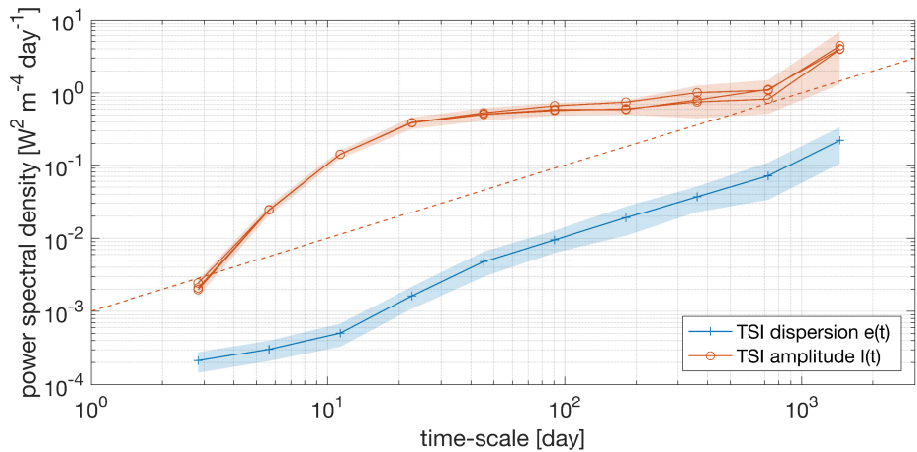}
\caption{In red: Power spectral density of the TSI. The three curves correspond to ACRIM3, VIRGO, and TIM. In blue: Power spectral density of the residual error $e(t)$. Confidence intervals correspond to a 1--$\sigma$ uncertainty. The spectral density is estimated by using a discrete wavelet transform with 4th-order Daubechies wavelets. The dashed line represents a $1/f$ scaling, with $f=1/$timescale.}
\label{fig_scaling}
\end{center}
\end{figure}

To determine how the uncertainty scales with frequency, we consider the dispersion between the instruments. Let $I_j(t)$ be the TSI from ACRIM3, VIRGO, and TIM, which are the three instruments that have the longest overlapping period (10.5 years). After centering each record by subtracting its time-average over the considered time-interval ($\tilde{I}_j(t) = I_j(t) - \langle I_j(t) \rangle_t$) we determine the residual error, defined as $e_j(t) = \tilde{I}_j(t) - \frac{1}{3} \sum_{i=1}^3 \tilde{I}_i(t)$. This error quantifies the discrepancy between the three instruments, and its power spectral density provides a frequency scaling appropriate for the three contributing TSI instruments, see Figure~\ref{fig_scaling}.

Interestingly, the power spectral density of the residual error scales almost as $1/f$, where $f$ is the frequency, and thus strongly departs from a white-noise assumption. The same scaling is observed for any combination of TSI instruments with overlapping observations, regardless of their duration, and thus is a robust result. $1/f$ noise, also known as flicker noise, arises in many contexts from shot noise in resistors to seismic oscillations near sea coasts, but, to the best of our knowledge, this is the first published report of flicker noise in solar-irradiance observations. If the residual error were instead dominated by linear trends, which would result in a $1/f^2$ scaling, stability could be defined in terms of TSI-error per unit time. This concept of stability is frequently used, for example, to estimate the net uncertainty $N$ years away simply by multiplying the current stability by $N$. Our findings suggest that this simple linear concept is not correct because the uncertainty is instead fundamentally timescale dependent. While net uncertainty does not increase as fast as linearly with time, it does have a long-term memory which is absent in the common white-noise assumption, for which one would have a frequency-independent scaling $1/f^0$. This result cautions the confidence by which most TSI models are extrapolated backward in time by relying on only a few decades of observations.


We are unable to determine how this $1/f$ scaling extends beyond decadal timescales. Assuming that it does at least up to the 40-year duration over which the TSI has been observed, our working hypothesis will be that each instrument is contaminated  by $1/f$ noise whose magnitude is set by the precision, as derived from Eq.~\ref{eq_AR_model}. This gives us a realistic and fully data-driven noise model by which all instruments can be meaningfully compared without resorting to subjective criteria. 

This model could be refined in several ways. For instance, including information from the instrument teams, such as increased precision uncertainty when there is documented evidence of a degradation in the observation conditions, could allow the inclusion of such a priori knowledge in the relative weighting of the data. The short-term uncertainties of ACRIM1, for example, are greater between November 1980 and April 1984 when the attitude control of the SMM satellite was degraded. Our following analyses, however, currently include no such refinements.


\section{Multiscale Decomposition}

A composite could in principle be built simply by doing a daily weighted-average of all available observations. This, however, would introduce artifacts with amplitude jumps  occurring whenever the number of observing instruments changes. To overcome this problem, we average both in time and scale-wise. First we decompose each TSI record $I(t) = \sum_k I(a_k,t)$ into multiple records that contain information at specific timescales $a_k$. The records from different instruments are then averaged scale-by-scale before we recombine them into one single composite. The wavelet transform is ideally suited for this. 

We require a wavelet transform that is redundant in time, translation-invariant (to be able to assign a precise time tag to each value of the wavelet transform) and orthogonal, so that different scales can be processed independently. The pyramidal wavelet transform with Gaussian kernels \citep{mallat08} fulfills these conditions.  

Wavelet transforms, however, require regularly-sampled records with no gaps. To overcome this, we first extrapolate all records over the full timespan from 17 November 1978 to 31 December 2015, then fill in all missing values, apply the wavelet transform, and finally, for each scale $a$, discard the wavelet transform $I(a,t_j)$ at those times $t_j$ for which observations are missing. The missing values are computed by expectation-maximization \citep{ddw11}. This approach makes no assumptions regarding solar variability except for the TSI records having high coherence. In particular, we do not impose any solar-cycle amplitude or trend.

After applying this expectation-maximization method, we end up with $N=8$ records (one for each instrument) of daily values that cover the same complete timespan. By bootstrapping, we find the 1--$\sigma$ uncertainty on the reconstructed values (for data gaps that are up to three months long) to be comparable to that of the observations. This gives us high confidence that the method does indeed offer a good approximation of the TSI. 

When computing the wavelet transform in the vicinity of a data gap, the latter will inevitably affect the value of the transform. What could be viewed as a weakness of the wavelet transform then actually becomes one of its main strengths. Let $I(a,t_i)$ be the wavelet transform at a given time $t_i$ with the nearest gap at time $t_g$. The transform will be influenced by the synthetic data from that gap only if it is located within a cone of influence, defined as $|t_i-t_g| < a$. Conversely, the wavelet transform at a data gap $I(a,t_g)$ will be influenced by nearby observations only if these are close enough. For ERBE, for example, which generally makes one measurement every 14 days, the wavelet transform associated with timescales $< 14$ days are primarily determined by extrapolated values of the TSI (i.e. on observations from other instruments, when available), whereas slower variations mainly reflect the observations made by ERBE and thus are much less affected by its numerous gaps.

This persistence of the wavelet transform in the vicinity of observations allows us to bridge data gaps in a natural way. In particular, it overcomes the aforementioned problem with the 2-year interruption between ACRIM1 and ACRIM2. To ensure that the method gives precedence to observations when there are nearby data gaps, we let the weight given to the wavelet transform drop off exponentially with a timescale-dependent decay time $a$ when moving away from the nearest observation (see below). In doing so, we allow the wavelet transform at a given time to include some information from nearby synthetic observations while severely restricting its impact to the cone of influence within which the wavelet transform is highly persistent.

The largest timescale is merely the average value of the TSI, which we chose to be the average of the TIM and the PREMOS absolute values.  

To summarize, we use a maximum-likelihood approach in which the composite is a weighted average of all the observations with weights that are classically defined as the inverse-squared uncertainty. This averaging is performed on a scale-by-scale basis. The main steps are:

\begin{enumerate}
\item Fill in all missing values by expectation-maximization and flag them. 
\item For each record $j = \{1, 2, \ldots, N \}$, estimate the precision $\sigma_j(t)$. If desired, include additional information to increase the precision manually during times when instrumental effects are known to affect the record.
\item Compute the scale-dependent uncertainty $\sigma_j(a,t)$ by extrapolating it with a $1/f$ model for the noise.
\item For each record, estimate the wavelet transform $I_j(a,t)$ at scales $a = \{2, 4, 8, \ldots 2^{N_s}\}$. For the largest scale, replace the wavelet transform by the time-average.
\item For each scale $a$, define the composite as a weighted average
\begin{equation}
I_{comp}(a,t) = \frac{\sum_{k=1}^N I_k(a,t) \; w_k(a,t)}{\sum_{k=1}^N w_k(a,t)}
\end{equation}
in which the scale- and time-dependent weights $w$ are defined as:
\begin{equation}
w_k(a,t) = \left\{ 
\begin{array}{ll}
\sigma_k^{-2}(a,t)  & 
\textrm{if instrument $k$ is} \\
 & \textrm{observing on day t} \\
\sigma_k^{-2}(a,t) e^{-2|T_k(t)|/a}  & 
\textrm{if instrument $k$ is not} \\
 & \textrm{observing on day t}
\end{array}
\right.
\end{equation}
wherein $T_k(t)$ is the temporal distance to the nearest observation for instrument $k$. These weights are illustrated in Appendix 2.
\item Apply the inverse wavelet transform to obtain $I_{comp}(t)$. 
\item Estimate the uncertainty of $I_{comp}(t)$ by using a Monte-Carlo approach in which ${>1000}$ composites are generated with additive noise as described by the noise model. 
\end{enumerate}


\section{The Composite}

\begin{figure}[!ht]
\begin{center}
\noindent\includegraphics[width=0.7\textwidth]{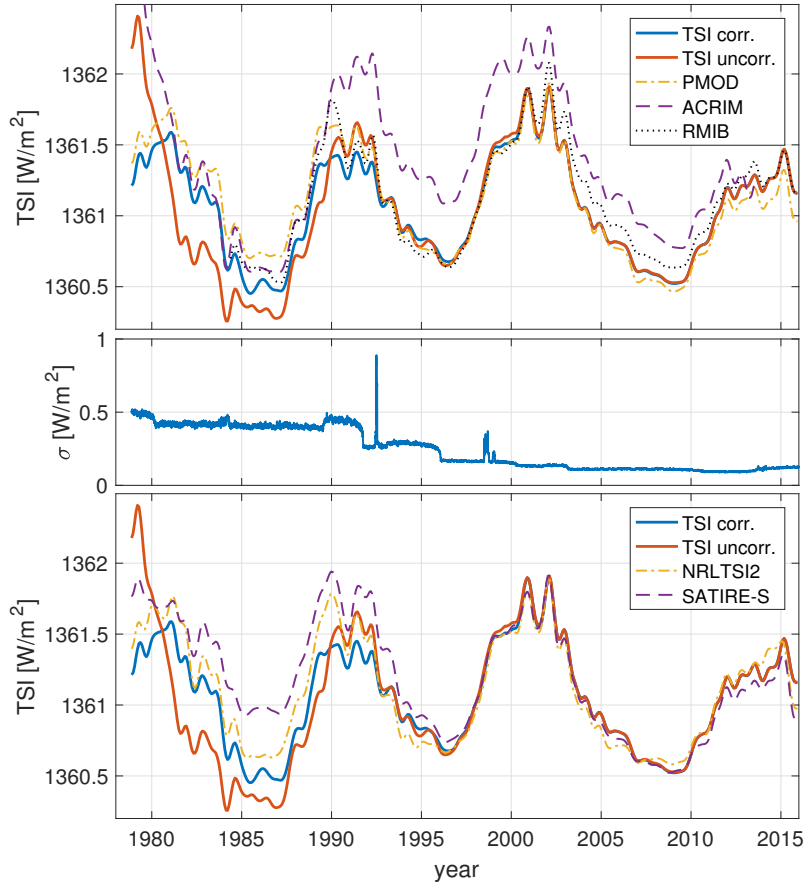}
\caption{Composite TSI obtained from corrected and uncorrected original records, and their comparison with existing observation-based composites (upper plot) and TSI models (lower plot). The middle plot shows the time-dependent uncertainty (standard deviation) of the composite based on our $1/f$ noise model, with daily resolution. All the other time series show 6-month averages. Note that the same vertical scale is used for all the plots.}
\label{fig_composite}
\end{center}
\end{figure}

Figure~\ref{fig_composite}  presents the resulting TSI composite and compares it to the three primary existing measurement-based composites (ACRIM, PMOD, RMIB) as well as to model reconstructions by SATIRE-S \citep{yeo14} and NRLTSI2 \citep{coddington16}. We provide two versions of the composite. The first version is based on the original TSI records as provided by the instrument teams without any correction or rescaling. However, there is evidence that some of the older instruments suffer from uncorrected artifacts. One such example is a likely early signal increase in HF, ACRIM1, and ERBE. This affects all radiometers except the TIM but is better corrected in most instruments since those earlier three. \citet{froehlich06} has corrected several of these records, which we incorporate in the second (so-called corrected) version of our TSI composite. 


There is currently growing consensus that Fr\"ohlich's corrections are a justifiable improvement to the original data. Let us nevertheless stress that neither of our two composites is definitive; our prime objective here is merely to reveal how corrections made to the original data affect the composite. 

Figure~\ref{fig_composite} shows that these corrections affect the composite most notably prior to 1985, but additionally cause a limited impact up to 1993. These early-era differences mainly stem from the initial on-orbit degradations in HF and ACRIM1, with the latter differences due to similar degradation that may affect ERBE over longer timescales. Both are within the 1--$\sigma$ confidence interval, but the final composite's uncertainties could be reduced if it were known what corrections should be applied.

The decreasing uncertainty seen in Figure~\ref{fig_composite} mainly reflects an improvement in precision from the newer instruments, and to a lesser degree the larger number of simultaneous observations. Large abrupt peaks occur whenever there are lengthy measurement-gaps.

Though our composites are not definitive, comparisons with other reconstructions are illuminating. The agreement of the corrected version of the composite with the PMOD composite is excellent, whereas the large upward trend exhibited by the ACRIM composite exceeds the uncertainty. Regarding models, we find a closer agreement with NRLTSI2 than with SATIRE-S, whose downward trend between successive solar minima is larger than supported by the observations. Note, however, that the difference between the two models is still within the estimated confidence interval. 

Of particular interest for solar and climate studies is the multi-decadal trend exhibited by the TSI during successive solar minima. Both composites show an increase between the minima of 1986 and 1996, followed by a decrease. Given the instrument stabilities, only the downward trend between 1996 and 2009 is statistically significant, see Appendix 3. The long-term memory effect of the $1/f$ uncertainty is an important ingredient here, for it generates more variability between the different solar minima than a classical, but unrealistic white noise model.


\section{Conclusion}

Our new approach brings several major improvements to the longstanding problem of merging of multiple observations into a single, fully-traceable composite: 1) the method uses all the available data instead of daisy-chaining or choosing a single long-duration instrument as the primary reference (the backbone method); 2) it relies on a data-driven noise model whose uncertainties are estimated in a systematic way without pre-conceived bias; 3) data gaps are bridged in a natural way thanks to the multi-scale nature of the method.  

The new TSI composite we obtain is not definitive because the original data still require some community-endorsed corrections. Future versions will also incorporate additional information, such as implementing greater uncertainties during periods of known instrument issues. Meanwhile, we find our composites in closer agreement with that from PMOD than those from ACRIM or RMIB, and similarly closer to the NRLTSI2 model than to the SATIRE-S model. Possible trends between solar minima are too weak to be statistically significant except for the downward trend between 1996 and 2009. Finally, we find the power spectral density of the uncertainty in comparisons between instruments to scale with frequency as $1/f$. As a consequence, the concept of a single-valued stability that is independent of timescale is not appropriate, which precludes extrapolating uncertainties forward or backward in time merely via simple linearly-growing values.  

Our approach can be readily extended to other types of data. We are presently applying it to spectrally-resolved solar irradiance data. One obvious, but mathematically-demanding improvement, is to move from a maximum-likelihood approach to a Bayesian one \citep[e.g.][]{tingley12}. This would provide a more natural way of merging observations that scale differently to each other, such as the MgII core-to-wing index or the sunspot number record.   

Most importantly, our approach decouples the statistical problem (\textit{What is the best way of constructing the composite?}) from the scientific one (\textit{What prior information goes into the correction of the original data sets?}). Eventually, the only means by which the user should be able to influence the composite's outcome is via estimates of the initial uncertainties, and not by adjusting the TSI records themselves. We consider this decoupling as a vital condition for obtaining an unbiased TSI composite.


\subsection*{Acknowledgments}
We acknowledge the principal investigators of the individual TSI instruments for making their valuable data available. TD and MS received funding from the European Community's Seventh Framework Programme (FP7-SPACE-2012-2) under the Grant Agreement 313188 (SOLID, \url{http://projects.pmodwrc.ch/solid/}). We thank the International Space Science Institute (ISSI, Bern) for hosting the team on the making of the new TSI composite. The authors declare that they have no conflict of interest. The  provisional TSI composite is available for download at http://www.issibern.ch/teams/solarirradiance and at the website of SOLID.




\section*{Appendix 1. Precision: order of the autoregressive model}
\label{appendix1}

We use an autoregressive model (Eq. 1 in the main text) to describe the variability of the the Total Solar Irradiance (TSI) $I(t)$ by means of a linear time-invariant model. A considerable amount of literature has been devoted to such models, see for example \citep{percival93,ljung97}. The model order $p$ is a compromise between the goodness of fit and a penalty for overfitting. Various information theoretic criteria have been developed for estimating $p$. For a broadband process such as the TSI, a simple approach consists of plotting the magnitude of the residual error $\epsilon(t)$ (namely, our precision $\sigma_I$) versus $p$, and determining when it stops decreasing significantly.

\begin{figure}[!ht]
\begin{center}
\noindent\includegraphics[width=0.8\textwidth]{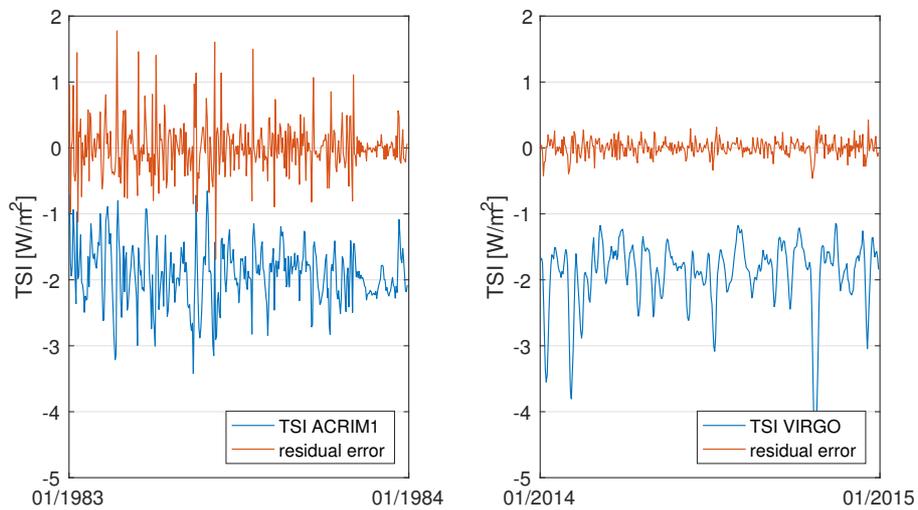}
\caption{Two excerpts of the residual error for models of order ${p=5}$, for ACRIM1 (left) and VIRGO (right). The TSI is shown on the same scale, but translated vertically. Solar activity levels, as indicated by sunspot numbers, were comparable for both time periods.}
\label{fig_tsi_resid}
\end{center}
\end{figure}

For early-generation instruments that suffer from large noise levels, the precision barely drops with the order $p$, and a first-order model suffices, indicating that the instrument precision is comparable to the actual TSI variability on daily timescales. For more recent instruments that have less noise, larger orders are required. Using data from these instruments, we observe no significant reduction in the precision for $p > 5$, and thus select $p=5$  for all datasets. Figure~\ref{fig_tsi_resid} illustrates the residual error for one old and one recent instrument, using this model order.

There exist alternative methods for estimating the precision \citep{ddw16} and we find them to be in close agreement with the values obtained from the autoregressive model approach. 

\begin{figure}[!ht]
\begin{center}
\noindent\includegraphics[width=0.7\textwidth]{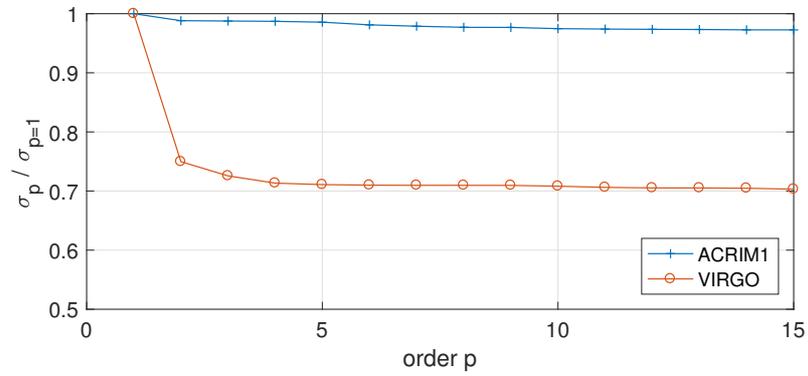}
\caption{Evolution of the precision with  model order $p$ for two datasets: ACRIM1, which has a large uncertainty, and VIRGO. Both curves are normalized to their maximum value for clarity.}
\label{fig_tsi_AR}
\end{center}
\end{figure}

\section*{Appendix 2. Time-dependent weights}
\label{appendix2}

\begin{figure}[!ht]
\begin{center}
\noindent\includegraphics[width=0.9\textwidth]{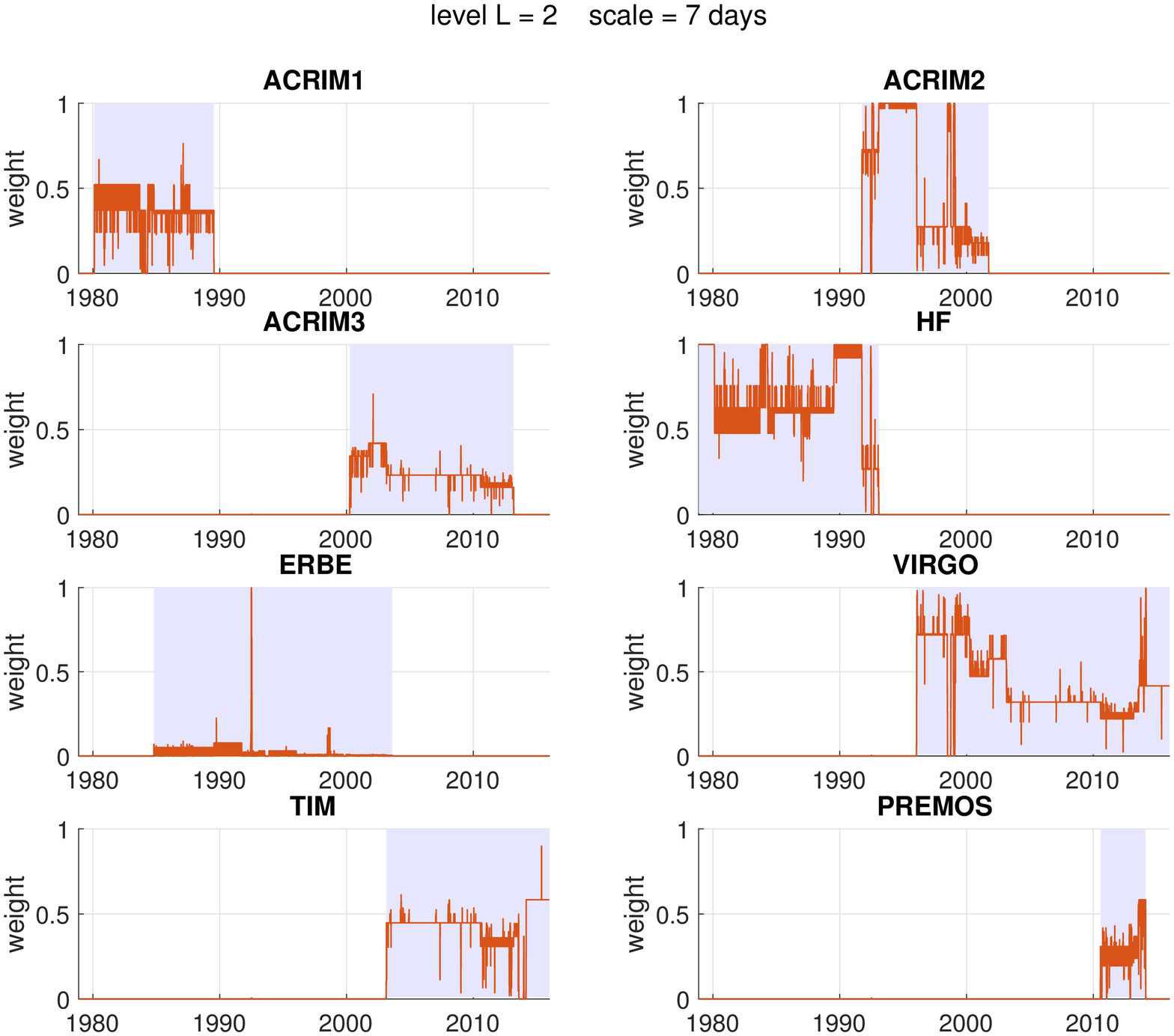}
\caption{Time-dependent weight determined from each instrument for a characteristic scale of 7 days (Level 2 of the wavelet transform). The shaded area represents the time interval during which the instrument was operating.}
\label{fig_tsi_L2}
\end{center}
\end{figure}

\begin{figure}[!ht]
\begin{center}
\noindent\includegraphics[width=0.9\textwidth]{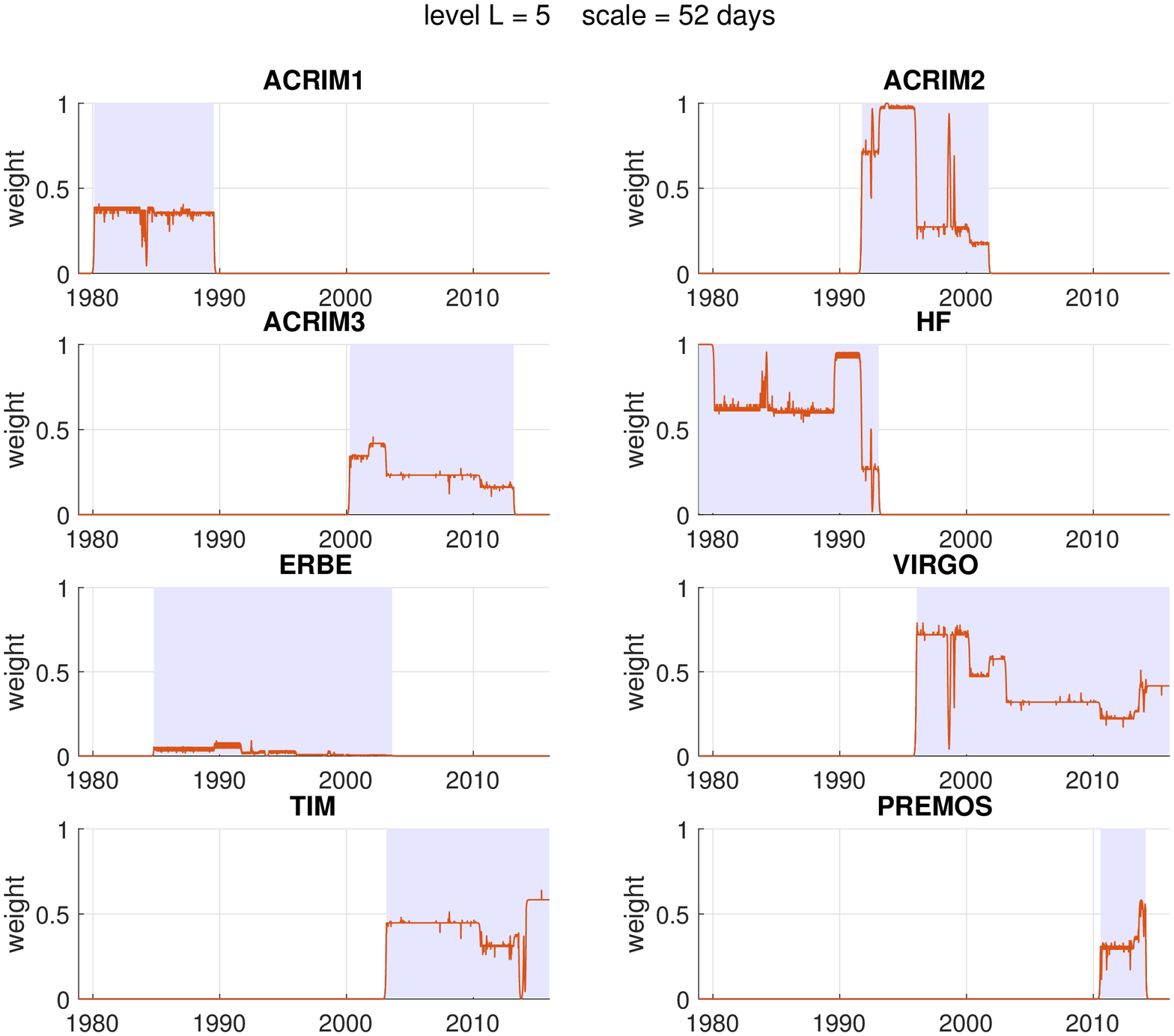}
\caption{Time-dependent weight determined from each instrument for a characteristic scale of 52 days (Level 5 of the wavelet transform). The shaded area represents the time interval during which the instrument was operating.}
\label{fig_tsi_L5}
\end{center}
\end{figure}

\begin{figure}[!ht]
\begin{center}
\noindent\includegraphics[width=0.9\textwidth]{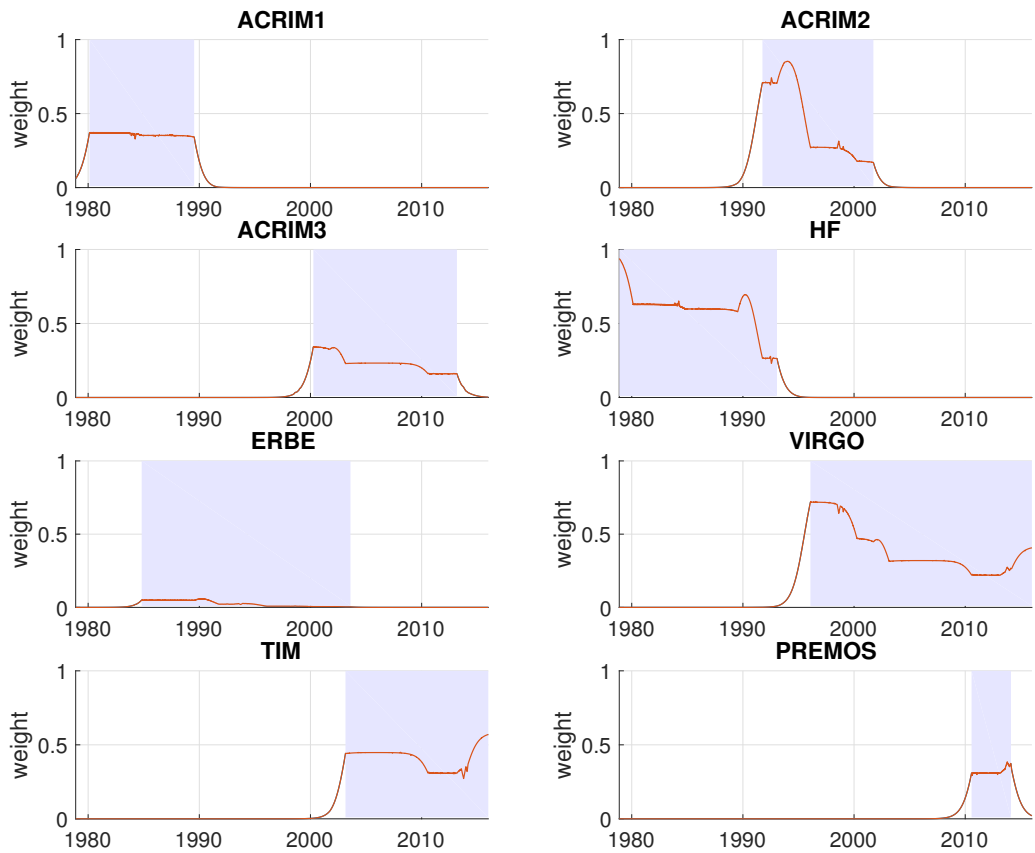}
\caption{Time-dependent weight determined from each instrument for a characteristic scale of 832 days (Level 9 of the wavelet transform). The shaded area represents the time interval during which the instrument was operating.}
\label{fig_tsi_L9}
\end{center}
\end{figure}

Figures~\ref{fig_tsi_L2} to \ref{fig_tsi_L9} illustrate the time-dependent weights for three timescales. The sum of the weights from all instruments adds up to one at any single time within each timescale. The exponential decay of the weights when moving away from an observation is best observed in Figure~\ref{fig_tsi_L9}, demonstrating the degree to which an instrument can influence the model even when not directly making observations and also the influence of that instrument on the weightings of other instruments.  Instruments having higher uncertainties, such as ERBE, have lower weightings.

\clearpage

\section*{Appendix 3. Uncertainty on long-term variations}
\label{appendix3}

Long-term variations in the TSI are often estimated by comparing averages made at successive solar minima. Using the uncertainties quantified by our composite, we can not only compare these averages but also assess how significant they are. 

Here, we determine the mean TSI by averaging its value over a one-year interval that is centered on each solar minimum. We first use a Monte-Carlo approach to build a large ensemble of composites (here, $N=50000$) and, subsequently, for each we compute the average at solar minimum. Finally we consider the differences between these averages: $\textrm{TSI}_{1996}-\textrm{TSI}_{1986}$, $\textrm{TSI}_{2009} - \textrm{TSI}_{1986}$ and $\textrm{TSI}_{2009} - \textrm{TSI}_{1996}$.  

\begin{figure}[!ht]
\begin{center}
\noindent\includegraphics[width=0.8\textwidth]{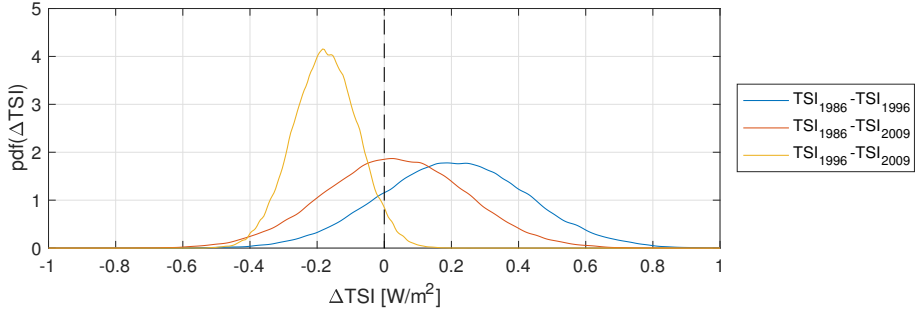}
\caption{Probability distribution functions (obtained by Gaussian kernel density estimation) of the difference $\Delta$TSI between different solar minima using an ensemble of ${N=50000}$ composites with a $1/f$ scaling of the uncertainty.}
\label{fig_tsi_test}
\end{center}
\end{figure}

The probability distributions of these three differences are displayed in Figure~\ref{fig_tsi_test}; they show considerable dispersion, with both positive and negative values. The difference between the most recent two solar minima has the least dispersion because of the better instrument precision.  To express the significance of the trends we simply consider the probability of obtaining a positive TSI difference, see Table~\ref{diffTSI}. With this criterion, the probability of having a downward trend between 1996 and 2009 is 0.96. This trend is statistically significant if we consider a significance level of $\alpha = 5\%$.

\begin{table}[ht]
\begin{center}
\begin{tabular}{|c|c|c|c|}
\hline
$\Delta$TSI & $\langle \Delta\textrm{TSI} \rangle$ & $\sigma_{\Delta\textrm{TSI}}$ & Probability   \\ 
 & [W/m$^2$] & [W/m$^2$] & $\Delta\textrm{TSI}> 0$ \\ \hline
$\textrm{TSI}_{1996} - \textrm{TSI}_{1986}$ & \ 0.21 	& 0.22 	& 0.82 \\
$\textrm{TSI}_{2009} - \textrm{TSI}_{1986}$ & \ 0.03 	& 0.21 	& 0.55 \\
$\textrm{TSI}_{2009} - \textrm{TSI}_{1996}$ & -0.17 & 0.10 	& 0.04 \\ \hline
\end{tabular}
\caption{Change in TSI between solar minima. $\langle \Delta\textrm{TSI} \rangle$ is the average of the difference and $\sigma_{\Delta\textrm{TSI}}$ is its standard deviation. The last column gives the probability of an upward trend. \label{diffTSI}}
\end{center}
\end{table}

If we now consider a more classical but unrealistic white noise model (with the same precision), then the Monte-Carlo modeled scatter in $\Delta$TSI drops by about 25 and all three trends become highly significant. This illustrates the importance of having a realistic noise model that takes into account long-term effects.


\end{document}